# RF photonic delay lines using space-division multiplexing

S. Garcia and I. Gasulla

ITEAM Research Institute, Universitat Politècnica de València, 46022 Valencia, Spain

## ABSTRACT

We review our last work on dispersion-engineered heterogeneous multicore fiber links designed to act as tunable true time delay lines for radiofrequency signals. This approach allows the realization of fiber-distributed signal processing in the context of fiber-wireless communications, providing both radiofrequency access distribution and signal processing in the same fiber medium. We show how to design trench-assisted heterogeneous multicore fibers to fulfil the requirements for sampled true time delay line operation while assuring a low level of crosstalk, bend sensitivity and tolerance to possible fabrication errors. The performance of the designed radiofrequency photonic delay lines is evaluated in the context of tunable microwave signal filtering and optical beamforming for phased array antennas.

## 1. INTRODUCTION

The application of Space-Division Multiplexing (SDM) technologies has been welcomed as a promising approach to overcome the upcoming capacity crunch of conventional singlemode fiber communications, [1]. In essence, SDM aims to increase the transmission capacity over a fixed bandwidth by increasing the number of light paths that are transmitted within a single optical fiber. For the last few years, different fiber possibilities have been investigated including multicore fibers (MCFs) [2-5], few-mode fibers (FMFs) [6] or a combination of both [7]. In the case of MCFs, most of the research has focused on homogeneous multicore fibers where all the cores share the same propagation characteristics. To increase the number of cores per cross-sectional area, heterogeneous fiber compositions were proposed in 2009, [3]. The cores in a heterogeneous MCF are arranged so that the crosstalk between any pair of neighboring cores is reduced as the phase matching condition is prevented, reducing as a consequence the core pitch of the fiber. Initial designs of heterogeneous 19-core fibers obtained a core pitch reduction down to 23 µm in a 125-µm cladding diameter by accommodating three different core compositions [3]. The introduction of trench-assisted configurations led in addition to larger core multiplicities as, for instance, the 30-core fiber designed with four different types of cores in a cladding diameter of 228 µm, [4]; and the 37-core fiber fabricated by Sasaki et al. using three types of cores arranged in a 248-um cladding diameter, [5].

Despite MCFs were initially envisioned in the context of core and metro optical networks, they can be applied to a wide range of areas including radio access networks and multiple antenna connectivity [8-10], radiofrequency (RF) signal processing [11] and multi-parameter optical fiber sensing [12]. In particular, RF signal processing applications, such as tunable microwave signal filtering, radio beamsteering in phased array antennas, arbitrary waveform generation and optoelectronic oscillation, can benefit from the use of MCFs in terms of compactness and weight, as well as performance stability and versatility. Nowadays, the trend is to dedicate a given system (either a fiber-based device or a photonic integrated circuit) to process the signal and a separate optical fiber link to perform its distribution. We envision a new approach where we implement the required signal processing functionalities while we distribute the RF signal to the end user (wireless base station, indoor antenna, etc.). This leads to the idea of "fiber-distributed signal processing", a concept with great potential regarding future converged fiber-wireless telecommunications networks, as those envisioned for 5G systems and the Internet of Things, [9].

Photonic true time delay lines (TTDLs) are at the heart of RF signal processing, [11,13]. This basic building block provides a frequency independent group delay for the modulating signal within a given frequency range. Figure 1 illustrates the general idea that underlies our proposal for sampled delay line operation built upon a heterogeneous multicore fiber. A set of replicas of the modulated signal travels along the MCF so that every replica is transmitted through a different core with a given group velocity. The goal is to obtain at the output of the fiber link a set of time-delayed samples where the differential delay $\Delta\tau = \tau_{n+1} - \tau_n$ is constant between any pair of adjacent samples. To achieve this, we must design the heterogeneous MCF to behave as a group-index-variable delay line where each core features an independent group delay with a different group delay dependence on the optical wavelength, as we will see in this manuscript.

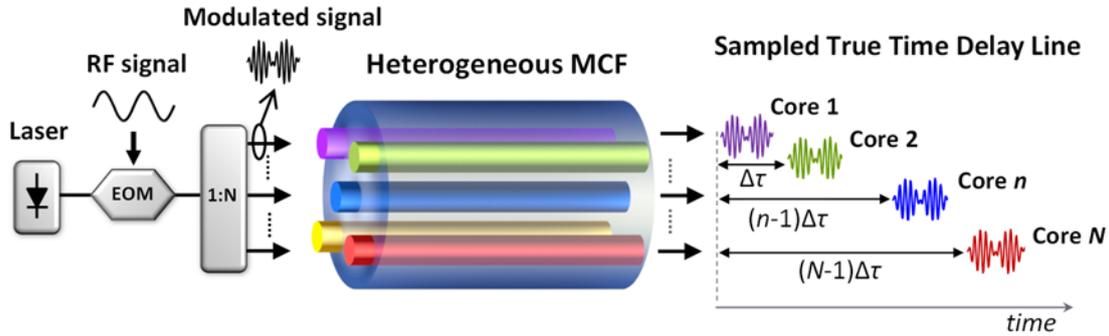

Figure 1. Sampled true time delay line built upon a heterogeneous multicore fiber for 1-dimensional operation.

We identify this configuration as 1D (1-dimensional) time delay operation since the different samples are given by the spatial diversity (i.e., core diversity) behavior of the MCF. However, we must take into account that this approach offers as well 2D (2-dimensional) operation if we incorporate the optical wavelength diversity provided, for instance, by an array of lasers emitting at different optical wavelengths, [14-16]. As illustrated in Fig. 2, if we operate using the optical wavelength diversity domain, the basic differential delay $\Delta\tau$ results from the propagation difference experienced by two adjacent wavelengths that travel in a given core. If we use the spatial diversity instead, the basic differential delay is created from the propagation difference experienced by the same optical wavelength when it propagates through two adjacent cores. The combination of both dimensions, space and optical wavelength, provides a wide range of true time delay possibilities (different sets of samples and basic differential delays) within the same single optical fiber.

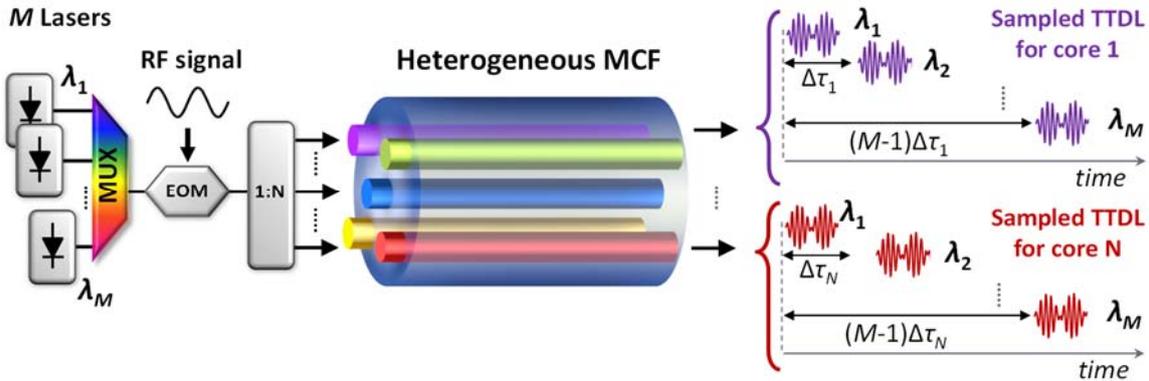

Figure 2. Sampled true time delay line built upon a heterogeneous multicore fiber for 2-dimensional operation.

We review in this paper our latest work on the area of radiofrequency signal processing using heterogeneous multicore fibers. In section 2, we address how to design dispersion-engineered heterogeneous MCFs to act as tunable fiber-distributed true time delay lines. We evaluate in section 3 the main sources of degradation that could compromise the performance of the designed photonic delay lines. As a proof of concept, we evaluate in section 4 the performance of the designed MCFs as distributed signal processing elements when they are applied to two typical Microwave Photonics (MWP) functionalities: tunable microwave signal filtering and optical beamforming networks for phased array antennas. Finally, section 5 provides a summary wrap-up, some conclusions and directions for future work.

## 2. SAMPLED DELAY LINE OPERATION OVER A HETEROGENEOUS MCF LINK

Up to date, most of the research activity in heterogeneous MCFs is focused on high-capacity digital communications, [3-5,7]. While keeping equal propagation characteristics (in particular the value of the chromatic dispersion parameter $D$) in all the cores is important in digital SDM applications, for the implementation of tunable TTDLs it is important to have different values at a fixed wavelength. Therefore, a new procedure for designing heterogeneous MCFs has to be developed where, while keeping the advantages in terms of crosstalk and immunity to bends, one has the flexibility of tailoring the chromatic dispersion parameter of each individual core. The design of heterogeneous MCFs to operate as group-index-variable delay lines implies that each core features an independent group delay with a linear dependence on the optical wavelength, [14-16] The group delay $\tau_{g,n}(\lambda)$ of a particular core $n$ can be expended in its 2nd-order Taylor series' around the anchor wavelength $\lambda_0$ as

$$\tau_{g,n}(\lambda) = \tau_{g,n}(\lambda_0) + D_n(\lambda - \lambda_0) + \frac{1}{2} S_n (\lambda - \lambda_0)^2 \quad , \tag{1}$$

where $D_n$ is the chromatic dispersion parameter and $S_n$ the dispersion slope of core $n$ at $\lambda_0$, [16] For proper TTDL operability, $D_n$ must increase linearly with the core number $n$ while keeping a linear behavior of the group delay with the optical wavelength. As shown in Fig. 3, if we force all cores to share the same group delay value $\tau_{g,0}$ at $\lambda_0$, we can control the basic differential delay between samples, $\Delta_{n,n+1} = \tau_{g,n+1}(\lambda) - \tau_{g,n}(\lambda)$, to provide continuous tunability from 0 up to tens (or even hundreds) of ps/km. This gives us the possibility of implementing fiber-distributed signal processing links with lengths up to a few kilometers. As Eq. (1) shows, to reduce the quadratic wavelength-dependent term, we must analyze higher-order dispersion effects meticulously. This group-index-variable delay line can work on two different regimes whether we exploit the spatial or the optical wavelength diversities [14,16]. The basic differential delay $\Delta\tau$ depends actually on the operation regime and, as consequence, we must address each regime individually for proper dispersion optimization.

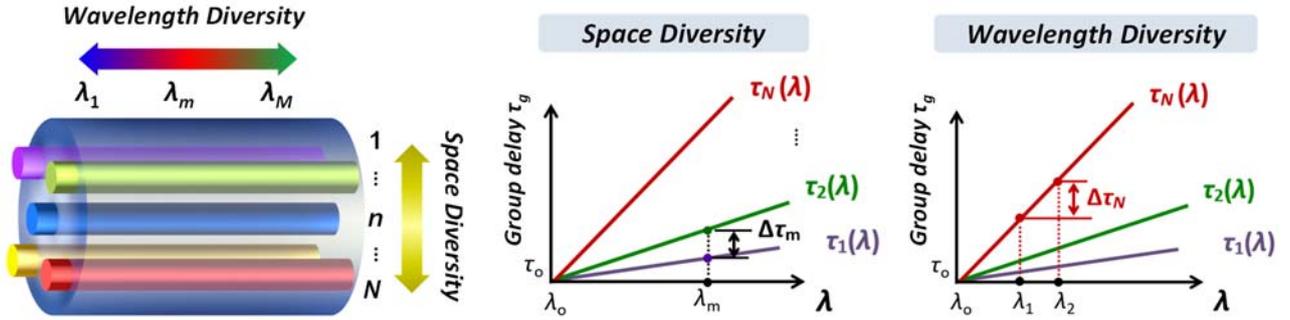

Figure 3. Spectral group delays for the N cores of the MCF showing both spatial and optical wavelength diversities.

In the case of spatial diversity, the differential group delay $\Delta\tau_{n,n+1}$ is given by the propagation difference created between each pair of adjacent cores for a particular optical wavelength $\lambda_m$:

$$\Delta\tau_{n,n+1}(\lambda_m) = \Delta D(\lambda_m - \lambda_0) + \frac{S_{n+1} - S_n}{2}(\lambda_m - \lambda_0)^2 \quad , \tag{2}$$

where $\Delta D = D_{n+1} - D_n$ is the incremental dispersion parameter that is kept constant for every pair of cores. We see from Eq. (2), that the differential group delay is affected by a quadratic term that depends on the difference between the dispersion slopes of the cores involved. To minimize this undesired term, we must design the cores of the MCF such as to have the most similar dispersion slopes $S$ possible.

When the delay line operates in optical wavelength diversity, the differential group delay experienced between two contiguous wavelengths ($\lambda_m, \lambda_{m+1}$) in a particular core $n$ is given by:

$$\Delta\tau_n(\lambda_m, \lambda_{m+1}) = D_n \delta\lambda + S_n(\lambda_1 - \lambda_0)\delta\lambda + \frac{1}{2} S_n (2m-1)\delta\lambda^2 \quad , \tag{3}$$

where $\delta\lambda = \lambda_{m+1} - \lambda_m$ is the separation between the two adjacent optical sources, $\lambda_1$ the wavelength of the first optical source and $1 \leq m \leq M - 1$, being $M$ the total number of optical sources. In this case, the undesired variation on $\Delta\tau_n$ is characterized by both the linear ($\delta\lambda$) and the quadratic ($\delta\lambda^2$) dependence on the dispersion slope. To minimize the higher-order dispersion effect we must therefore minimize as much as possible the effect of the dispersion slope $S_n$ as compared to the chromatic dispersion $D_n$.

With the previous design requirements in view, we have proposed and evaluated several tunable TTDLs based on different designs of heterogeneous multicore fibers, [15-16]. There, the use of trench-assisted cores improves not only the intercore crosstalk robustness, but also offers more design versatility since more design parameters are involved. Figure 4(a) shows the schematic of a trench-assisted step-index refractive index profile, where $a_1$ is the core radius, $a_2$ is the core-to-trench distance, $w$ is the trench width, $\Delta_1$ is the core-to-cladding relative index difference, and $\Delta_2$ is the cladding-to-trench relative index difference. By a suitable selection of these parameters, it is possible to obtain the required linearly incremental chromatic dispersion values and the common group delay at the anchor wavelength. In [16], we found an optimum TTDL design in terms of both time delay tunability in a broad optical wavelength range (i.e., higher-order dispersion minimization) and crosstalk for a 7-core fiber with trench-assisted step-index core profiles. Figures 4(b-c) illustrate the group delay and chromatic dispersion parameter computed for each core as a function of the core radius. There, the filled circles represent the selected core radius values and the corresponding group delay (Fig. 4(b)) and chromatic dispersion (Fig. 4(c)) values. Table 1 gathers all the core design variables as well as the computed dispersion slopes and effective indices. By using the numerical software Fimmwave, we obtained a range of $D$ values from 14.75 up to 20.75 ps/(km·nm) with an incremental dispersion $\Delta D = 1$ ps/(km·nm). Table 1 gathers all the core design variables as well as the computed dispersion slopes, effective indices and effective areas. The cladding-to-trench relative index was fixed to $\Delta_2 \approx 1\%$ in all cores.

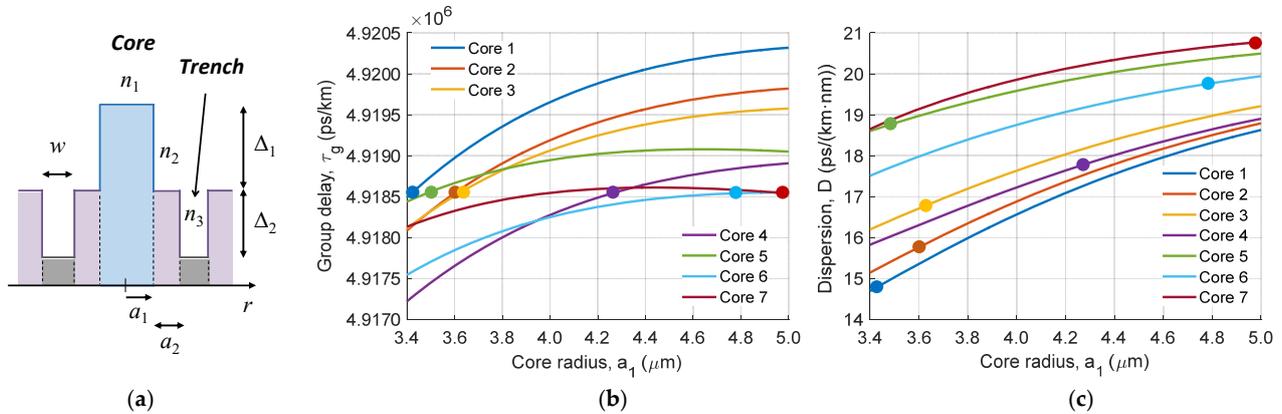

Figure 4. (a) Trench-assisted refractive index profile of the fiber cores; (b) Group delay $\tau_g$ and (c) chromatic dispersion $D$ of each core as a function of the core radius $a_1$. Each filled circle represents the selected core radius for the 7-sample TTDL design and the corresponding group delay and chromatic dispersion values.

Table 1. Core design parameters and computed fiber characteristics of the designed heterogeneous MCF.

| Core | $a_1$ (μm) | $a_2$ (μm) | $w$ (μm) | $\Delta_1$ (%) | $D$ (ps/(km·nm)) | $S$ (ps/(km·nm$^2$)) | $n_{eff}$ |
|---|---|---|---|---|---|---|---|
| 1 | 3.42 | 5.48 | 3.02 | 0.3864 | 14.75 | 0.065 | 1.4534 |
| 2 | 3.60 | 5.03 | 2.61 | 0.3762 | 15.75 | 0.064 | 1.4535 |
| 3 | 3.62 | 4.35 | 3.32 | 0.3690 | 16.75 | 0.065 | 1.4534 |
| 4 | 4.26 | 4.92 | 4.67 | 0.3588 | 17.75 | 0.064 | 1.4539 |
| 5 | 3.49 | 2.81 | 5.41 | 0.3476 | 18.75 | 0.064 | 1.4529 |
| 6 | 4.79 | 3.35 | 3.32 | 0.3435 | 19.75 | 0.064 | 1.4540 |
| 7 | 4.98 | 2.42 | 4.05 | 0.3333 | 20.75 | 0.064 | 1.4540 |

Figure 5 (a) shows the cross section of the designed MCF, where the cores are placed in a hexagonal disposition with a 35-μm core pitch and 125-μm cladding diameter. Figure 5(b) shows the computed group delay for each core as a function of the operation optical wavelength. We observe that the requirements for TTDL operability are fulfilled, that is: (1) incremental values of the group delays with a constant basic differential delay between cores; (2) common group delay at the anchor wavelength $\lambda_0 = 1550$ nm; and (3) linearly incremental group delay slopes. This allows us to implement tunable TTDLs that can be used as the basic element to perform multiple MWP applications, as we will show in section 4. But, first, we will review the main sources of degradation that could comprise the performance of the heterogeneous MCF as a tunable true time delay line.

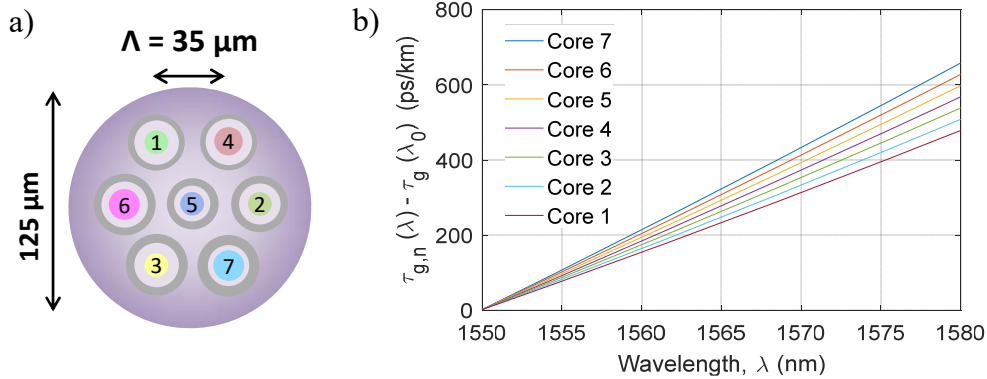

Figure 5. (a) Cross-sectional view of the designed heterogeneous MCF; (b) Computed core group delays versus optical wavelength.

## 3. POSSIBLE DEGRADATIONS

### 3.1 Crosstalk

Heterogeneous MCFs were conceived to reduce the intercore crosstalk and thus increase the core density as the phase matching condition between adjacent cores is prevented. Actually, one of the major detrimental effects that can degrade the performance of heterogeneous MCFs is the crosstalk dependence on the phase-matching condition between adjacent cores when the fiber is bent, [17-19]. A bent MCF can be described as the corresponding straight fiber with equivalent core refractive indices, [20],

$$n_{eq,n}(r, \theta, R_b) = n_n(r, \theta)(1 + r \cos \theta / R_b), \quad (4)$$

where $R_b$ is the bending radius, $n_n$ the core refractive index of the straight fiber and $(r, \theta)$ the local polar coordinates of each core, being $\theta$ the angle from the radial direction of the bend. The threshold bending radius $R_{pk}$ is the maximum curvature that can induce a phase matching between any pair of adjacent cores and is then expressed as, [20],

$$R_{pk} = \Lambda \cdot n_{eff,n} / |n_{eff,n} - n_{eff,m}|, \quad (5)$$

where $\Lambda$ is the core pitch and $n_{eff,n}$ is the effective index of core $n$. To minimize the threshold bending radius as much as possible, we designed the core refractive index profiles and placed the cores inside the fiber cross section as to maximize the difference in the effective indices of adjacent cores. The location of the cores can be seen in Fig. 5(a). We simulated using FIMMPROP numerical software the worst-case crosstalk between adjacent cores versus the bending radius for a 10-km MCF link. We see from Fig. 6 that the threshold bending radius is located at 103 mm, as expected from Eq. (5), and the crosstalk level above the phase matching region is kept below -90 dB.

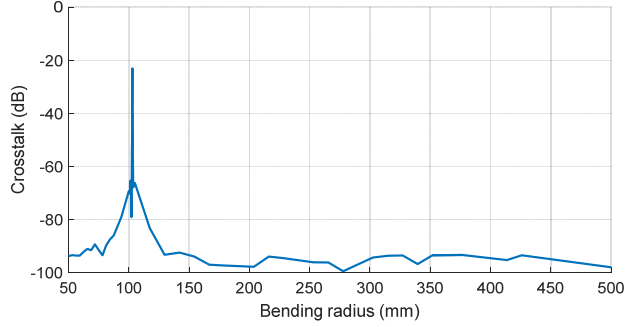

Figure 6. Computed worst-case crosstalk as a function of the bending radius.

### 3.2 Curvature and twisting effects

Apart from the well-known influence of bending and twisting effects on the inter-core crosstalk in MCFs, the delay characteristics of the fiber can also be affected by the fiber curvatures. From Eq. (4), we see that the worst case variation on the group delay and dispersion characteristics due to bends is when $\theta = 0$. Figure 7 represents the simulation of the worst-case variation (i.e., for $\theta = 0$) on the (a) group delay and (b) dispersion of each core as a function of the bending radius $R_b$. We see in both cases that all the outer cores behave similarly when they are located in the radial direction of the curvature. For the particular case of the group delay, we observe an important degradation of hundreds of ps/km even for large bending radii. In contrast, Fig. 7(b) shows that the dispersion variation due to fiber curvatures can be neglected even for small bending radii.

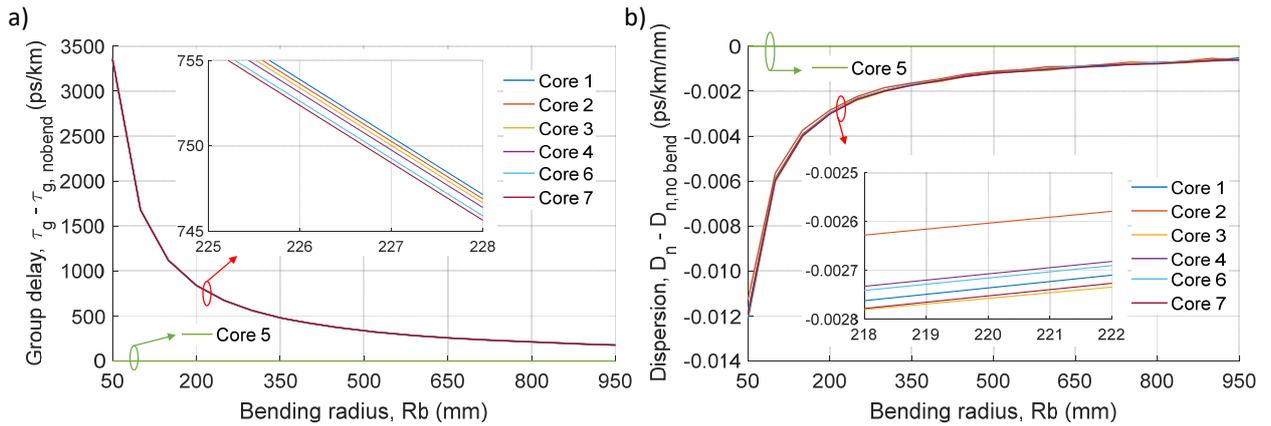

Figure 7. Computed worst-case variation of the (a) group delay and (b) chromatic dispersion of each fiber core as a function of the bending radius $R_b$.

While we have seen that the group delay variation due to fiber curvatures would have important consequences on the time delay performance of the true time delay line, the twisting effect plays an important role here to minimize this effect. Figure 8 shows the dependence of the group delay of each core as a function of the angle $\theta$ for a bending radius of $R_b = 50$ mm. We see again that all the outer cores behave similarly, following a cosine function as expected from Eq. (4). A fiber twist can be understood as a rotation of the fiber cross-section, so that the $\theta$-coordinate of a given core will change according to this rotation. This means that the equivalent index that characterizes a given core when the fiber is bent (see Eq. (4)) equals the straight index if a $2\pi$ continuous-rate twist and a constant bending radius are applied. Thus, if the fiber is coiled in a spool, the twist-induced perturbations will contribute to cancel the expected bend-induced variations on the core group delays.

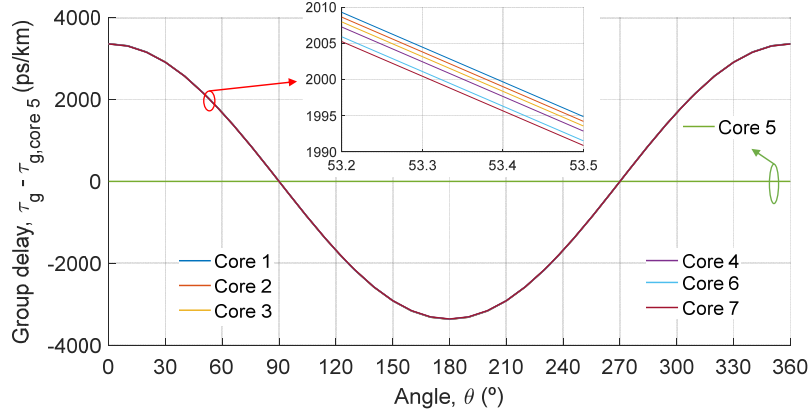

Figure 8. Computed group delay variation as a function of the angle $\theta$ from the radial direction of the bend for each core at a bending radius of 50 mm.

### 3.3 Fabrication tolerances

We have evaluated numerically possible variations of the group delay and chromatic dispersion values of the fiber cores due to expected fabrication tolerances. The most critical parameters here are the core radii. In particular, the current tolerances of MCF fabrication processes are in the order of ±0.1 μm for the radii. Figure 9(a) represents the computed group delays of the cores as a function of the optical wavelength when these fabrication tolerances are considered. Here, we set a random perturbation with uniform distribution between ±0.1 μm for each core radii. We see that the performance of the delay line is highly degraded due to the inclusion of these tolerances, mostly due to the error on the group delay. The group delay degradation can be avoided by properly compensating the variation of the core group delays with external delay lines. The chromatic dispersion variation is less critical, and its robustness could be furtherly improved if the incremental dispersion $\Delta D$ between cores is enlarged. Figure 7(b) shows the computed core group delays as a function of the optical wavelength when the error on the group delays is compensated. The required compensation of the group delay would also induce a variation on the chromatic dispersion of the fiber, but this additional value would be several orders of magnitude lower than the dispersion of the corresponding core. Comparing this response with Fig. 5(b), we can see that the effect of the fabrication tolerances can be mostly solved just by externally compensating the errors on the group delay.

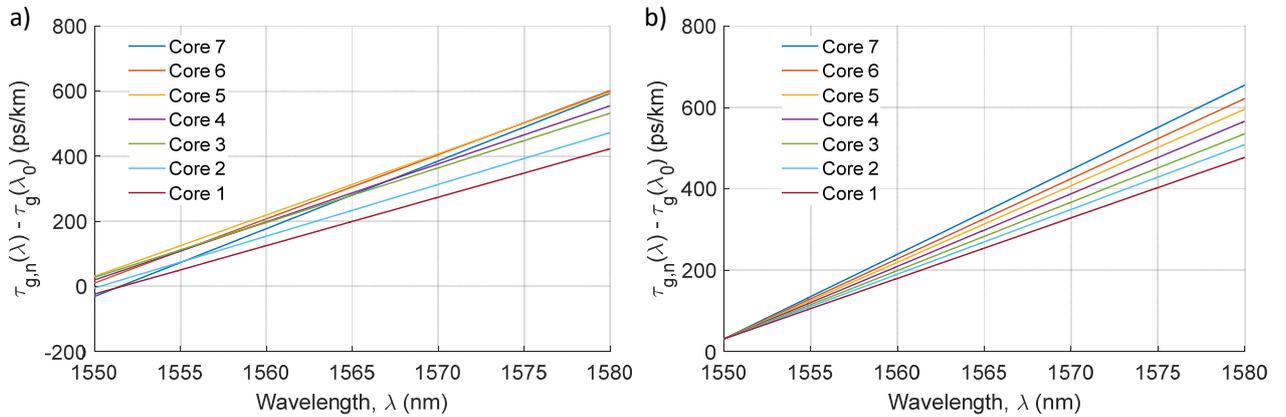

Figure 9. Computed core group delays versus optical wavelength when a random perturbation due to fabrication tolerances is applied to the group delays and chromatic dispersions of the cores. (a) Without delay compensation; (b) with delay compensation.

# 4. MICROWAVE PHOTONICS APPLICATIONS

The developed delay lines serve as a compact and energy efficient solution to implement a variety of signal processing functionalities that will be demanded in future fiber-wireless communications networks and subsystems, [11]. As a proof of concept, we evaluate here the performance of the designed MCFs as distributed signal processing elements when they are applied to two typical MWP functionalities: tunable microwave signal filtering and optical beamforming networks for phased array antennas. For simplicity, we focus only on exploiting the space-diversity regime.

## 3.3 Tunable microwave signal filtering

A frequency filtering effect over RF signals results from combining and collectively photodetecting (with a single receiver) the delayed signal samples coming from the TTDL output. This incoherent Finite Impulse Response filter is characterized by a transfer function $H(f)$ that is given by, [3]:

$$H(f) = \sum_{n=0}^{N-1} a_n e^{-jn2\pi f \Delta \tau}, \tag{6}$$

where $a_n$ is the weight (amplitude and phase) corresponding to the $n^{th}$ sample and $f$ the RF frequency. The frequency period or Free Spectral Range (FSR) of the filter is given by $FSR = 1/\Delta\tau$, where $\Delta\tau$ is the basic differential delay as defined in section 2. If we exploit the spatial diversity domain of our MCF-based TTDL, as shown in the scheme depicted in Fig. 10(a), this effect occurs when the $N$ output signals from the $N$ cores are photodetected in a single receiver. We have computed the corresponding RF frequency responses when we employ the 7-core fiber analyzed in the previous section, whose design and propagation parameters are gathered in Table 1. Figure 10(b) represents the computed electrical transfer functions of the microwave signal filter implemented with a 10-km link of this fiber. As expected, this delay line allows to continuously tune the FSR of the filter by changing the operation wavelength of the laser. For instance, we see how changing the operation wavelength from 1560 up to 1575 nm reduces the FSR from 10 down to 4 GHz without any significant degradation due to higher-order dispersion effects.

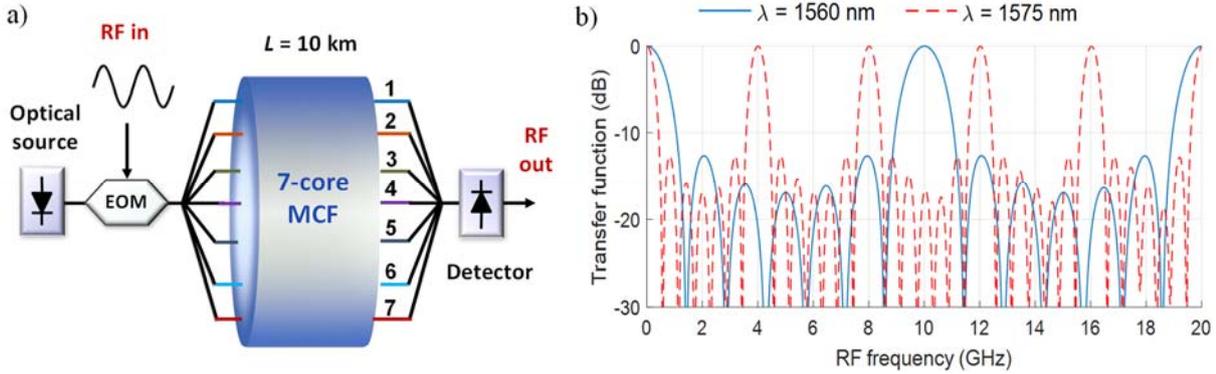

Figure 10. (a) Schematic view of the 7-core-fiber-based microwave signal filter; (b) Computed electrical transfer function as a function of the RF frequency for an operation wavelength of 1560 nm (Blue solid lines) and 1575 nm (Red dashed lines).

## 3.4 Optical beamforming for phased array antennas

Optical beamforming networks are implemented using a similar configuration as in microwave signal filtering, with the particularity that each sample is individually photodetected and, then, feeds one of the radiating elements that conformed the phased array antenna. Figure 11(a) shows the scheme of a MCF-based optical beamforming network when exploiting space diversity for a 10-km link. In the case of 1D architectures, the normalized angular far-field pattern of the radiated electric field, or array factor $AF(\theta)$, is given by:

$$AF(\theta) = \sum_{n=0}^{N-1} a_n e^{-j2\pi n v (\Delta \tau - d_x \sin(\theta)/c)}, \tag{7}$$

where $\theta$ is the far field angular coordinate, $v$ is the optical frequency ($v = c/\lambda_0$) and $d_x$ is the spacing between adjacent radiating elements. From Eq. (7), the direction $\theta_0$ of maximum radiated energy can be adjusted by tuning the basic differential delay since $\Delta\tau = d_x \sin(\theta_0) / c$. Figure 11(b) shows the computed array factor as a function of the beam pointing angle (in degrees) in both polar coordinates (left) and decibels (right) for $d_x$ = 3 cm and a 5-GHz RF signal. Again, we show the TTDL tunability when varying the input optical wavelength. We appreciate how changing the operation wavelength from 1560 up to 1575 nm steers the radio beam direction from 180º down to 90°.

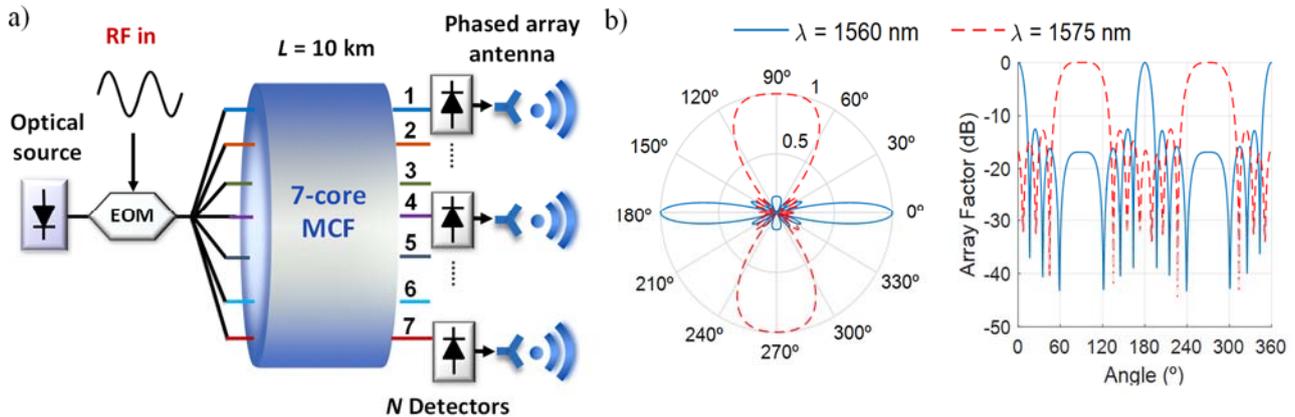

Figure 11. (a) Schematic view of the 7-core-fiber-based optical beamforming network for a phased array antenna; (b) Computed array factor as a function of the beam-pointing angle for an operation wavelength of 1560 nm (Blue solid lines) and 1575 nm (Red dashed lines).

## 5. CONCLUSIONS

We have shown that the introduction of the space dimension into the implementation of optical true time delay lines results in compact and versatile fiber-distributed signal processing solutions for microwave and millimeter wave signals. In particular, we have reviewed how to design trench-assisted heterogeneous multicore fibers that are optimized in terms of higher-order dispersion and crosstalk to operate as a broadband tunable delay lines. This approach will be key in next-generation fiber-wireless communications systems and access networks, where the radiofrequency signal will be processed while it is been distributed, for instance, from a central office to the end user. The reported SDM technology will benefit these scenarios in terms of compactness as compared to a set of parallel singlemode fibers, performance stability against mechanical or environmental conditions and operation versatility offered by the simultaneous use of the spatial- and wavelength-diversity domains. Future research will be accomplished in the framework of the European Consolidator Grant InnoSpace awarded by the European Research Council, including among others the fabrication of the proposeddispersion-engineered heterogeneous MCFs and the subsequent experimental demonstration of microwave photonics signal processing applications.

**ACKNOWLEDGMENTS**

This research was supported by the ERC Consolidator Grant 724663, the Spanish MINECO Projects TEC2014-60378-C2-1-R and TEC2016-80150-R, the Spanish scholarships MINECO BES-2015-073359 for S. García and the Spanish MINECO Ramon y Cajal program RYC-2014-16247 for I. Gasulla.